# NUCLEAR EMULSION FILM DETECTORS FOR PROTON RADIOGRAPHY: DESIGN AND TEST OF THE FIRST PROTOTYPE

S. BRACCINI[†], A. EREDITATO, I. KRESLO, U. MOSER, C. PISTILLO, S. STUDER

*Albert Einstein Center for Fundamental Physics,*
*Laboratory for High Energy Physics (LHEP), University of Bern,*
*Sidlerstrasse 5, CH-3012 Bern, Switzerland*

P. SCAMPOLI

*Dipartimento di Scienze Fisiche, Università di Napoli Federico II,*
*Complesso Universitario di Monte S. Angelo, I-80126, Napoli, Italy*
*and*
*Department of Radiation Oncology, Inselspital, University of Bern,*
*Murtenstrasse 35, CH-3010 Bern, Switzerland*

Proton therapy is nowadays becoming a wide spread clinical practice in cancer therapy and sophisticated treatment planning systems are routinely used to exploit at best the ballistic properties of charged particles. The information on the quality of the beams and the range of the protons is a key issue for the optimization of the treatment. For this purpose, proton radiography can be used in proton therapy to obtain direct information on the range of the protons, on the average density of the tissues for treatment planning optimization and to perform imaging with negligible dose to the patient. We propose an innovative method based on nuclear emulsion film detectors for proton radiography, a technique in which images are obtained by measuring the position and the residual range of protons passing through the patient's body. Nuclear emulsion films interleaved with tissue equivalent absorbers can be fruitfully used to reconstruct proton tracks with very high precision. The first prototype of a nuclear emulsion based detector has been conceived, constructed and tested with a therapeutic proton beam at PSI. The scanning of the emulsions has been performed at LHEP in Bern, where a fully automated microscopic scanning technology has been developed for the OPERA experiment on neutrino oscillations. After track reconstruction, the first promising experimental results have been obtained by imaging a simple phantom made of PMMA with a step of 1 cm. A second phantom with five 5 x 5 $mm^2$ section aluminum rods located at different distances and embedded in a PMMA structure has been also imaged. Further investigations are in progress to improve the resolution and to image more sophisticated phantoms.

---

[†] Corresponding author - E-mail: Saverio.Braccini@cern.ch.

## 1. Introduction

Proton therapy is nowadays becoming a wide spread clinical practice in cancer therapy and many hospital based centers are under construction or planned[1]. The high precision of this innovative technique in radiation therapy allows a very high local control of the pathology with minimal secondary effects. This is particularly advantageous in the treatment of tumors located near organs at risk and in pediatric radiation oncology where the irradiation of healthy tissues may lead to severe permanent consequences on the quality of life of the patients and to the induction of secondary cancers.

To exploit at best the ballistic properties of charged particles, sophisticated treatment planning systems are routinely used to calculate the dose given to the target and to the nearby organs to be spared. To accomplish this task, imaging techniques represent more and more a crucial issue and the precise knowledge of the behavior of the beam inside the patient is of paramount importance for the optimization of the treatment in proton therapy.

## 2. Proton radiography with nuclear emulsions

The possibility to use high-energy protons to obtain medical images of the patient's body – the so-called proton radiography – is an interesting research issue since many years[2].

This imaging methodology is based on the use of protons having a fixed energy, larger than the one used for therapy, thus penetrating the patient's body. By measuring the residual range, proton radiography allows obtaining images directly proportional to the average density of the traversed material, thus reducing range uncertainties in proton therapy treatment planning, usually based on computed tomography images to which correction factors are applied. Moreover, the dose given to the patients is order of magnitudes less with respect to ordinary X-ray radiography since every single proton passing through the patient's body brings valuable information.

Important results in proton radiography have been obtained at PSI using scintillator based detectors[3] and several research groups in Europe and US are exploring this imaging modality using sophisticated and expensive semiconductor based detectors[4].

We propose[5] an innovative method based on nuclear emulsion film detectors. This technique relies on high precision charged particle tracking inside the emulsions, as routinely performed in neutrino physics experiments. As presented in Fig. 1, nuclear emulsion films interleaved with tissue equivalent absorbers can be fruitfully used to reconstruct proton tracks with very high precision to obtain images through the measurement of the position and the residual range of protons passing through the patient's body.

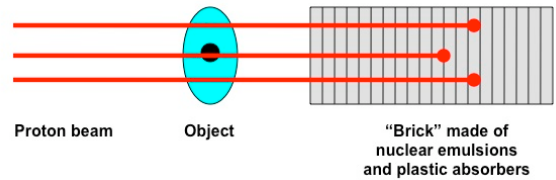

Figure 1 - Schematic principle of a detector for proton radiography based on nuclear emulsion films.

The proposed technique has good potentialities since a detector of this kind is very cheap and easy to install and remove. These characteristics are surely valuable in a clinical environment where routine daily treatment has absolute priority. Another possible application is the characterization of proton beams for new proton therapy clinical facilities.

## 3. Construction and test of the first prototype

To study the feasibility of this new application of nuclear emulsions, simulations based GEANT3 and SRIM has been performed in order to image two different phantoms. The first 'step' phantom is made of PMMA with thicknesses of 3 and 4 cm, thus presenting a step of 1 cm. The second 'rod' phantom is made of PMMA with a total thickness of 4.5 cm in which five 5 x 5 mm$^2$ section aluminum rods are embedded in different positions.

Several configurations of the stack – dubbed 'brick' - made of nuclear emulsions and polystyrene layers have been considered in order to optimize the resolution on the measurement of the range of the protons and to minimize the number of emulsions to be employed. A 138 MeV proton beam, uniformly spread on a 10 x 10 cm$^2$ surface, has been considered. The emulsion films used for this work consist of 2 layers of 44 μm thick sensitive emulsion coated on both sides of a 205 μm thick triacetate base.

Following the results of the simulations, two identical bricks have been constructed. Each brick is composed of two parts. The first part is made of 30 emulsions, interleaved with 1.5 mm polystyrene absorber plates and allows the tracking of passing through protons using a minimal number of emulsions. The second part is composed by 40 emulsions interleaved with 0.5 mm polystyrene and allows a precise measurement of the range. The Bragg peak is located in the second part, where the spatial resolution is maximal.

The clinical beam of the Gantry 1 at PSI has been used for the first beam tests, as presented in Fig. 2 (left). In order to be able to reconstruct the proton tracks without saturating the emulsions, a maximum of $10^6$ protons on a 10 x 10 cm$^2$ surface are needed. This corresponds to an average dose to the patient smaller than $10^{-5}$ Gy. For this purpose a special low intensity beam was set-up by the PSI team and an accurate measurement of the intensity was performed using a coincidence scintillating telescope before performing the irradiation. The irradiation was performed with 138 MeV protons uniformly spread on the surface by means of spot scanning.

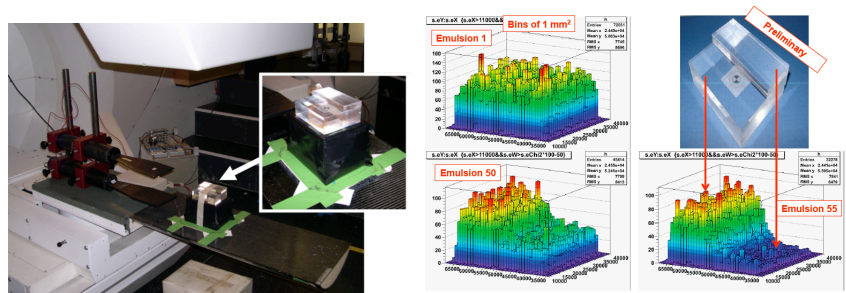

Figure 2 - First beam tests of proton radiography with nuclear emulsions at the Gantry1 at PSI. The brick with the 'step' phantom on the top is highlighted in the inset (left). The density of the mini-tracks is plotted for different emulsion sheets (right).

After the development, the scanning of the emulsions has been performed at LHEP in Bern, where a fully automated microscopic scanning technology has been developed for the OPERA experiment on neutrino oscillations.

Before full track reconstruction, mini-tracks contained only in one emulsion sheet have been searched for. As presented in Fig. 2 (right), in the case of the 'step' phantom, a clear decrease of the mini-track density has been observed in correspondence of the Bragg peaks produced by protons crossing 3 and 4 cm of PMMA, respectively. Emulsion 1, which is the nearest to the phantom, shows a proton density of about $10^6$ protons per 10 x 10 cm$^2$, as expected. The Bragg peak corresponding to protons passing through 4 cm of PMMA is located in correspondence of Emulsion 50, according to the simulations. Emulsion 55 is traversed only by protons passing through 3 cm of PMMA.

After full track reconstruction, the residual range is measured for all the proton tracks. For the 'step' phantom', a clear image of the 1 cm PMMA step is obtained, as shown in Fig. 3 (left). A preliminary estimation of the resolution of about 2.5 mm is obtained. For the 'rod' phantom, the measurement of the range as a function of the position shows the image of the five rods, as presented in Fig. 3 (right). The first peak has a better resolution and corresponds to the rod nearest to the brick. It is interesting to note that by measuring the average scattering angle only with Emulsion 1, the effect due to the five rods can clearly be put in evidence.

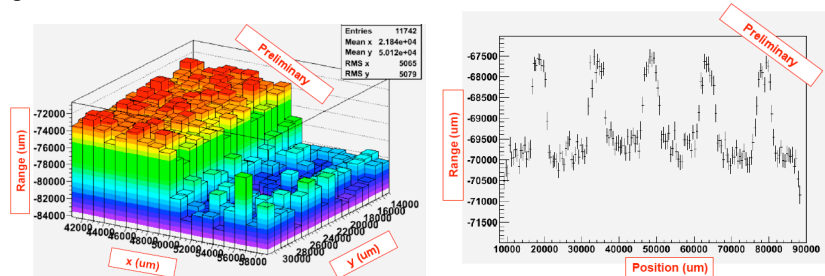

Figure 3 – Proton radiography image of the 'step' phantom (left). Projected proton radiography of the 'rod' phantom (right).

Following these first positive results, further investigations are in progress at LHEP to test new kind of nuclear emulsion films, improve the resolution and image more sophisticated phantoms.

**Acknowledgments**

The authors would like to acknowledge the Proton Therapy centre of the Paul Scherrer Institute (PSI) in Villigen (Switzerland) for setting-up and providing the proton beam.